\documentclass{aastex}          
\usepackage{spr-astr-addons}    

\begin{document}
%
\title{Hot subdwarfs in Resolved Binaries}

\shorttitle{Hot subdwarfs in Resolved Binaries}
\shortauthors{S.~J.~O'Toole}

\author{Simon J. O'Toole\altaffilmark{}} 
\affil{Anglo-Australian Observatory, PO Box 296, Epping NSW 1710, Australia}
\email{otoole@aao.gov.au} 


\begin{abstract}
In the last decade or so, there have been numerous searches
for hot subdwarfs in close binaries. There has been little to no
attention paid to wide binaries however. The advantages of
understanding these systems can be many. The stars can be assumed to
be coeval, which means they have common properties. The distance and
metallicity, for example, are both unknown for the subdwarf component,
but may be determinable for the secondary, allowing other properties
of the subdwarf to be estimated. With this in mind, we have started a
search for common proper motion pairs containing a hot subdwarf
component. We have uncovered several promising candidate systems,
which are presented here. 
\end{abstract}


\section{Introduction and Motivation}

The mass of a star is of fundamental importance to our understanding
of stellar evolution. While there are several robust methods for
determining the masses of stars on the Main Sequence, for more evolved
non-degenerate stars the
situation is much more difficult and many assumptions must be
made. Stars at the extreme blue end of  the horizontal branch - the
hot subdwarfs - fall into this category. 

Hot subdwarfs (sdB, sdO) are core helium-burning stars that are found
at the blue end of the horizontal  branch. They are found in all
Galactic stellar populations and are sufficiently numerous to account
for  the UV-upturn of early-type galaxies. While it is generally
accepted they will evolve directly into  white dwarfs and not return
to the giant phase, exactly how the stars arrived at this point in the
HR  diagram is still an open question. There are several possible
formation channels: binary evolution  through mass transfer and common
envelope ejection \citep[e.g.][]{Han2003};  delayed core helium
flashes in post-Red Giant Branch (RGB) evolution,
\citep[e.g.][]{Lanz2004}; the merger of two helium-core white dwarfs
\citep[e.g. ][]{SJ2000}; and non-core helium burning post-RGB
evolution \citep[e.g.][]{CC1993}. The  first three channels give rise
to a mass distribution peaked near 0.48\,$M_\odot$, while the latter
channel produces helium-core white dwarfs with masses closer to
$\sim$0.3\,$M_\odot$.

There are only a handful of useful parallax measurements of hot
subdwarfs because they were -- in general -- too faint to be observed
with Hipparcos \citep[see][]{Heber2002}. This in turn means they have very
few reliable mass determinations.  \citet{Heber1992} noted \emph{``Hot
subluminous stars in binary systems could provide an important tool
for checking the evolutionary scenarios ... since they possibly allow
stellar masses to be determined. Visual binaries and eclipsing
spectroscopic binaries are of utmost importance in this respect.''}
Indeed,
about half of  the sdBs reside in close binaries with white dwarf or
late-type Main Sequence companions. There are  several eclipsing
spectroscopic binaries containing the latter combination, however
these all also show  a reflection effect and therefore rely on our
incomplete knowledge of the impacts of stellar irradiation  \citep[for
example, see][]{Drechsel2001,Vuckovic2007}. 

There are some tremendous advantages of studying a hot subdwarf in a
resolved binary system. Firstly, the stars \emph{do not} interact, and
most likely have \emph{never} interacted. This means one does not have
to worry about the physics of binary interactions. Secondly, the
individual components can be studied independently, without the
problems and assumptions surrounding composite spectra. Lastly, Main
Sequence and Red Giant Branch stars are \emph{well understood}, at
least relative to hot subdwarfs.

We have therefore searched for common proper motion pairs and visual
binaries containing a hot subdwarf component. This is the focus of
the present work, where we present our most promising candidates and some
very preliminary results. The reader should note that there are
no definite or clear conclusions in the work presented here; this is
an exploratory pilot-type study, which is in part intended to inspire
others to explore alternative methods to determining the fundamental
parameters of these enigmatic objects.

\section{Visual binaries and Common Proper Motion companions}

With separations large enough to show no spectral contamination,
understanding visual binary and common proper motion systems
containing hot subdwarf components
will be more straightforward than other methods. It allow us to use
the spectroscopic parallax to determine the hot subdwarf's mass and
metallicity. (Note that while we use the terms ``spectroscopic
parallax'' and ``photometric parallax'' in this paper, they may be
considered as more related to a distance modulus than an angular
measurement.) The latter has previously been unknown in field hot
subdwarfs (unlike their stellar cluster counterparts), as atmospheric
diffusion leads to peculiar abundance patterns \citep[e.g.][]{OH2006}.

\begin{figure}[t]
\includegraphics[width=\columnwidth]{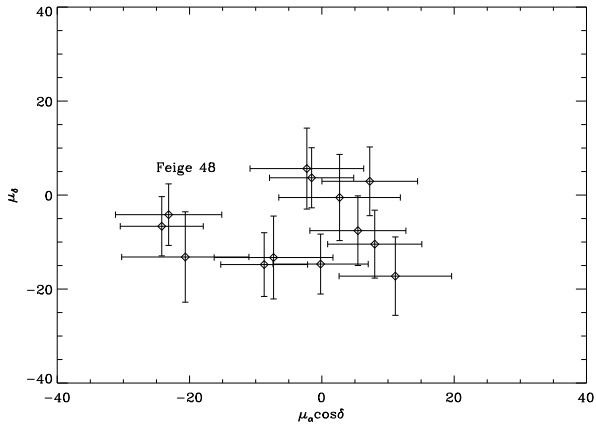}
\caption{SuperCosmos proper motions of stars in a field centered on
  the pulsating sdB Feige~48 (labelled)} 
\label{fig:feige48}
\end{figure}

And while pulsations detected in some sdB stars have allowed the
estimation of  masses of a handful of objects
\citep[e.g.][]{Charpinet2008}, they rely on single-star evolutionary
models where a mass range is assumed. Both mass and metallicity are
vital in understanding the evolution of both the RGB progenitor and
the hot subdwarf itself.

\subsection{Trawling through databases}

As a first step in this project, we have used the Subdwarf Database
\citep{Oestensen2006} as the source of known objects. Latter work will
use large databases such as the Sloan Digital Sky Survey Data Release
7\footnote{http://www.sdss.org/dr7/}. After identifying hot subdwarf
candidates, we then cross-correlated their co-ordinates with proper
motion and visual binary databases including, but not limited to:
The Washington Double Star Catalogue; SuperCosmos; UCAC2; NOMAD1; and
Hipparcos. Once pairs are identified, the goal of the project
is to determine parameters of companion star and the hot subdwarf. 

There are some limitations and difficulties with this approach to be
considered. The main concern is that hot subdwarfs are in general
quite distant and faint, which means that their proper motions are
often very small and/or not well constrained. Figure \ref{fig:feige48}
shows the proper motions for the known binary sdB pulsator Feige~48;
the star has a moderately high proper motion ($\sim -25$\,mas/yr), but
cannot be easily distinguished from the other stars in the field
around it. In the worst cases though, no proper motions have been
measured at all.

Once candidate pairs are selected, we follow two possible avenues to
determine their photometric parallax and therefore their distance:
empirical relations involving colours \citep{Hawley2002,Ivezic2008} or
relations using spectral indices \citep[e.g.][]{CR2002}. The combination of
two should lead to a more precise mass estimate. The photometric
parallax can then be combined with the spectroscopic parallax to give
the mass of the subdwarf. The spectroscopic parallax requires the
subdwarf's magnitude, effective temperature and surface gravity. The
mass is derived by matching the two parallaxes.

\subsection{The most promising candidates}

From our initial search described above, we have found these
promising candidates: BD+48\,1777; FBS 2254+373; PHL\,932; and
PG\,1618+563. Thomas Rauch has also discussed the possibility that
EC11481-2303 is also a member of a close pair (these
proceedings). Below we discuss the candidates in more detail.

\subsubsection{BD+48\,1777}

\begin{figure}
\includegraphics[width=\columnwidth]{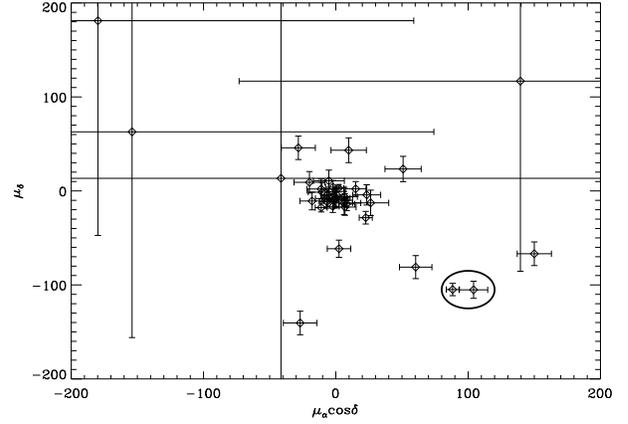}
\caption{SuperCosmos proper motions of stars in a field centered on
  the He-sdO BD+48\,1777. The sdO and its companion are circled} 
\label{fig:BDp48PM}
\end{figure}

This bright object is listed in the Luyten Double Star catalogue and
has a separation of 127 arcsec from its companion star. The hot
subdwarf has a Hipparcos parallax of 6.15$\pm$1.75\,milliarcsseconds
(mas), however the reliability of this value is uncertain. The
associated uncertainty is too high to meaningfully constrain the
star's distance and mass in any case. The corresponding Hipparcos
proper motions of $\mu_\alpha \cos\delta = 83\pm 2$\,mas/yr and
$\mu_\delta = -84\pm 1$\,mas/yr. The companion star is faint by
comparison -- it has SDSS photometry, with \emph{(u,g,r,i,z)} = (21.074,
18.512, 17.097, 16.138, 15.635) -- and has proper motions of
$\mu_\alpha \cos\delta =  80\pm 3$\,mas/yr and $\mu_\delta = -83\pm
3$\,mas/yr, almost an exact match. The proper motions from SuperCosmos
of stars with 150 arcseconds of BD+48\,1777 are shown in Figure
\ref{fig:BDp48PM}. The two stars (at $\mu_\delta\sim -100$\,mas/yr) are
clearly separated from the main group around the origin. Note that the
SuperCosmos proper motions shown in the figure are somewhat higher
than those from the more reliable and precise UCAC2 catalogue quoted
above.

\begin{figure}
\includegraphics[width=0.9\columnwidth]{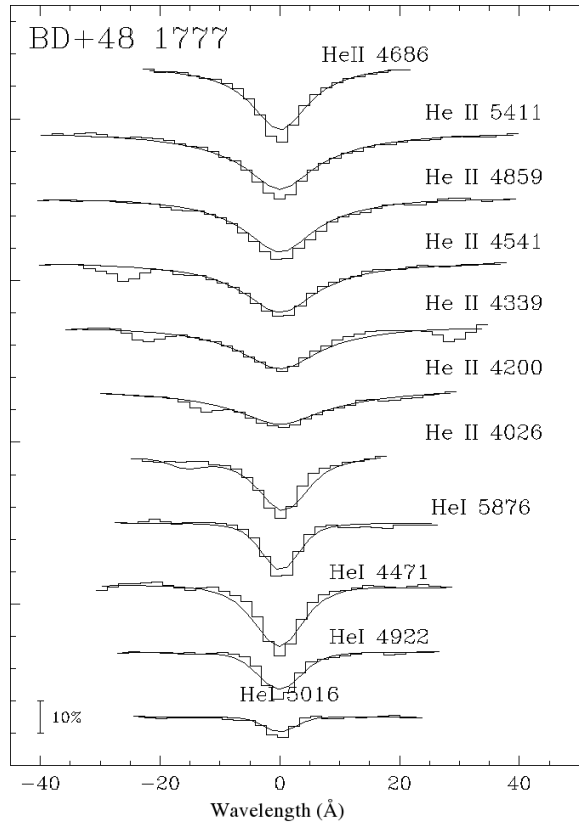}
\caption{Model fit of 6\,\AA\ spectrum of the He-sdO BD+48\,1777,
  kindly carried out by Uli Heber} 
\label{fig:BDp48fit}
\end{figure}

BD+48\,1777 is a reasonably well-studied object and we are fortunate
to have access to a spectrum, taken using CAFOS on the Calar Alto
2.2\,m with 6\,\AA\ resolution and covering wavelength range
3200--6350\,\AA. A model fit of this spectrum kindly carried out by Uli
Heber (and shown in Figure \ref{fig:BDp48fit}) using the NLTE model
grid of \citet{Stroeer2007}. The best-fit stellar parameters are given
in Table \ref{tab:BDp48}. The star is a ``garden-variety'' helium-rich
subdwarf O star and sits with the majority of these objects in the HR
diagram at around $\sim$45\,000\,K and $\log g\sim 5-6$.

\begin{table}
\begin{center}
\caption{Stellar parameters derived by Uli Heber for the He-sdO BD+48\,1777} 
\label{tab:BDp48}
\begin{tabular}{lc}
\tableline  
$T_{\mathrm{eff}}$ & 44664$\pm$147\,K \\
$\log g$ & 5.84$\pm$3 \\
log\,(He/H) & 2.69$\pm$0.00 \\
\tableline 
\end{tabular}
\end{center}
\end{table}

The extinction in the direction of the system is low ($\sim$0.04
mag). The SDSS colours of the companion star are consistent with it
being an M0 dwarf. If we use the empirical relations of
\citet{Hawley2002} we find a photometric parallax of $(i- M_i)
\approx 7.9$\,mag. The distance to the star is therefore
$\approx$\,380\,pc. This in turns leads to the absolute visual
magnitude of BD+48\,1777 of M$_V \approx$ 2.8\,mag, using $V =
10.7$ for the sdO star. We note that these values are very
preliminary. Unfortunately the spectrum we
have of the He-sdO is not flux calibrated, which makes mass determination
for the star currently difficult. 

Interestingly, the distance implies that the physical separation is
$\approx$48\,000\,AU, which is a separation usually found among young
binaries. It is expected that those binaries will not survive long
before becoming unbound gravitationally.

A spectrum of the M dwarf would lead to better parallax via a spectral
index method \citep[e.g.][]{CR2002,Reid1995}, while a flux-calibrated
spectrum of the He-sdO will lead to the first mass determination of
this class of star. The M dwarf spectrum would also allow the
determination of the progenitor system's metallicity, which would be
another first for \emph{any} hot subdwarf. Because it is so bright,
this potential makes BD+48\,1777 perhaps the most exciting candidate
system we have discovered so far.

\subsubsection{FBS\,2254+373}

\begin{figure}
\includegraphics[width=\columnwidth]{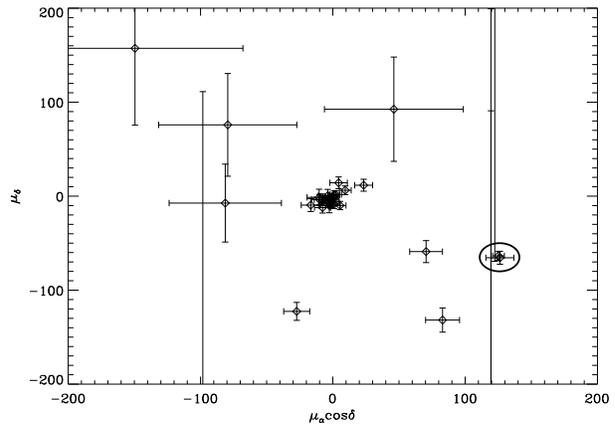}
\caption{SuperCosmos proper motions of stars in a field centered on
  FBS\,2254+373. The star and its companion are circled} 
\label{fig:FBS2254PM}
\end{figure}

This object was classified in the First Byurakan Survey as an sdOD star
-- in other words, a helium-rich subdwarf. Its proper motion is
145\,mas/yr. Unfortunately there is very little good quality
photometry of the star and it is relatively faint ($V\sim
16-17$\,mag), meaning that new photometric measurements are
required. The companion star, BD\,+36\,4962, has the same high
proper motion (see Figure \ref{fig:FBS2254PM}) and is very bright ($V
= 10.1$\,mag). Its 2MASS photometry suggests it is a late type star,
while its $B-V$ colours (0.99\,mag) are consistent with it being a K0
giant star. Somewhat frustratingly, neither star has any published
spectra, nor parameter determinations. The separation of the pair is
57 arcsec.

Note that a spectrum of FBS\,2254+373 kindly taken by Thomas Kupfer
after this paper was submitted has shown the object to be a DC
white dwarf rather than a hot subdwarf star. While this will make it
considerably more difficult to determine the mass, it remains
a very interesting object.

\subsubsection{PHL\,932}

This star, long believed to be the central star of a planetary nebulae
(although see \citet{Frew2010} who find that it is not), is the only
object in our sample that is well studied spectroscopically \emph{and}
that has a reliable trigonometric parallax \citep[3.36$\pm$0.62 mas --
][]{Harris2007}. The parallax places the star at a distance of
251-365\,pc, but when combined with the spectroscopic parameters from
e.g. \citet{Lisker2005}, the possible mass range is still large:
0.23-0.5\,$M_\odot$. This means that we cannot distinguish between a
star that has undergone core helium-burning (on the EHB) or not
(post-RGB). The latter has been assumed by e.g. \citet{Napiwotzki1999}
in previous discussion of the object.

The companion to PHL932 has been observed by the SDSS and has the same
proper motion ($\sim$38\,mas/yr from the NOMAD catalogue) and a very wide
separation (171 arcsec or $\sim$50,000\,AU at the approximate distance
to the system). Its SDSS and 2MASS colours are consistent with a
spectral type is M2-M3 when compared to Table 3 of
\citet{Hawley2002}. Using the empirical relation by the same authors,
we find $(i-M_i) = 7.2\pm 0.2$\,mag. This
constrains the star to a distance in the range 267-302\,pc, which in
turn leads to  a mass for PHL\,932 of 0.26-0.33\,$M_\odot$. This
almost certainly means that the object is \emph{not} an EHB star, but
rather an object that will evolved directly from the RGB to the white
dwarf cooling curve, and that will not ignite helium in its
core. There are still large uncertainties associated with this
determination, however. With a more precise measurement of its photometric
parallax, however, we will be able to constrain PHL932's evolutionary
history.

\subsubsection{PG\,1618+563}

This system is also very exciting. It consists of a
pulsating sdB star and an F3V star in a close ($\sim 3$\,arcsec)
binary. \citet{Silvotti2001} discovered the pulsations in the sdB, and
discussed the companion's spectral type and status, however did not go
any further. Both stars are relatively bright (sdB: 13.3, F3: 12.8),
so a
detailed study of this system is possible. With commonly used analysis
methods, the distance, metallicity and age of the F3V star should be
straightforward to measure, which in turn will lead to the mass of the
sdB star, independent of asteroseismological models. This system will
provide an excellent test of both evolution \emph{and} pulsation
models for sdB stars.

\section{Future Work}

We have presented the first results and motivation for an exciting new
method to determine the mass of hot subdwarf stars independent of
evolutionary models, which necessarily assume an atmospheric structure
and mass range.

There are several questions surrounding the systems presented here
that warrant further investigation. Some of these stars have very wide
physical separations; how could they stay gravitationally bound for so
long? Could they instead be two stars that were once in a closer
binary system and were then ``ejected'' after a mass-transfer phase?
Can we even describe them as binary systems or are they more similar
to co-moving groups, despite their age?

We need a large study of sdB proper motions using archival
and possibly new data. With the release of the UCAC3 catalogue, which
has larger sky coverage and higher precision proper motions (at least
in some cases) some progress is already being made in this
direction. The rise of survey telescopes (e.g. SkyMapper) may make it
more and more straightforward to obtain accurate proper motions in the
coming decade. On top of these surveys, the GAIA space mission will
deliver precise astrometry (to $\sim$0.3\,mas) of stars down to
V\,$\approx$\,20\,mag. Since space velocities are in three dimensions,
it will also be important to obtain relatively precise radial
velocities, allowing the status of the candidates presented here to be
confirmed.

We plan to carry out a detailed spectroscopic study of the companion
stars presented here, as well as several other candidates found in the
Sloan Digital Sky Survey database. We will use spectra to determine
photometric parallaxes of the late K and M dwarf
companions by measuring TiO-, CaH- and CaOH-band spectral indices
using the prescription of \citet{CR2002} and metallicities using the
calibration of \citet{WW2006}. We also plan to observe hot subdwarfs
with no available spectroscopy, and then measure their effective temperatures,
surface gravities and helium abundances using standard line-profile
fitting \citep[e.g. ][]{Lisker2005, Stroeer2007}. 

There are several late G-/early K-type companions among our
sample (not discussed here) and we will initally use the empirical
relationships of \citet{Ivezic2008}, before eventually obtaining high
resolution spectra and conducting a more detailed analysis of their
parameters.

The successful completion of this project will dramatically increase
the number of hot subdwarfs with accurate mass determinations
\emph{independent of evolution models}. We will also determine the
first accurate masses for helium-rich subdwarfs, which is a class of
stars that cannot be investigated using other methods, such as stellar
oscillations and asteroseismology. The masses will allow the first
robust tests of the proposed formation channels for some of these objects
(e.g. Justham, these proceedings).


\begin{thebibliography}{}
\bibitem[\protect\citeauthoryear{Castellani \& Castellani}{1993}]{CC1993}
Castellani \& Castellani, 1993, ApJ, 407, 649
\bibitem[\protect\citeauthoryear{Charpinet et al.}{2008}]{Charpinet2008}
Charpinet, S. et al. 2008, A\&A, 489, 377
\bibitem[\protect\citeauthoryear{Cruz \& Reid}{2002}]{CR2002}
Cruz, K. \& Reid, I.~N. 2002, AJ, 123, 2828
\bibitem[\protect\citeauthoryear{Drechsel et al.}{2001}]{Drechsel2001}
Drechsel, H. et al. 2001, A\&A, 378, 893 
\bibitem[\protect\citeauthoryear{Frew, Madsen \& O'Toole}{2010}]{Frew2010}
Frew, D., Madsen, G. \& O'Toole, S.~J., 2010, PASA, in press, astro-ph/0910.2078
\bibitem[\protect\citeauthoryear{Han et al.}{2003}]{Han2003}
Han, Z. et al. 2003, MNRAS, 341, 669
\bibitem[\protect\citeauthoryear{Harris et al.}{2007}]{Harris2007}
Harris, H. et al. 2007, AJ, 133, 631
\bibitem[\protect\citeauthoryear{Hawley et al.}{2002}]{Hawley2002}
Hawley, S. et al. 2002, AJ, 123, 3409
\bibitem[\protect\citeauthoryear{Heber}{1992}]{Heber1992}
Heber, U., 1992, Lectures in Physics, 401, 233
\bibitem[\protect\citeauthoryear{Heber et al.}{2002}]{Heber2002}
Heber, U., et al. 2002, A\&A, 383, 938
\bibitem[\protect\citeauthoryear{Ivezic et al.}{2008}]{Ivezic2008}
Ivezic et al., 2008, ApJ, 684, 287
\bibitem[\protect\citeauthoryear{Lisker et al.}{2005}]{Lisker2005}
Lisker, T., et al. 2005, A\&A, 430, 223
\bibitem[\protect\citeauthoryear{Lanz et al.}{2004}]{Lanz2004}
Lanz, T. et al. 2004, ApJ,  602, 342
\bibitem[\protect\citeauthoryear{Napiwotzki}{1999}]{Napiwotzki1999}
Napiwotzki, R., 1999, A\&A, 350, 101
\bibitem[\protect\citeauthoryear{\O stensen}{2006}]{Oestensen2006}
{\O}stensen, R., 2006, Baltic Astronomy
\bibitem[\protect\citeauthoryear{O'Toole \& Heber}{2006}]{OH2006}
O'Toole, S.~J. \& Heber, U., 2006,  A\&A, 452, 579
\bibitem[\protect\citeauthoryear{Reid et al.}{1995}]{Reid1995}
Reid, I.~N. et al. 1995, AJ, 110, 1838
\bibitem[\protect\citeauthoryear{Saio \& Jeffery}{2000}]{SJ2000}
Saio, H. \& Jeffery, C.~S. 2000, MNRAS, 313, 671
\bibitem[\protect\citeauthoryear{Silvotti et al.}{2000}]{Silvotti2001}
Silvotti, R., et al. 2000, A\&A, 359, 1068
\bibitem[\protect\citeauthoryear{Str{\"o}er et al.}{2007}]{Stroeer2007}
Str{\"o}er, A., et al. 2007,  A\&A, 462, 269
\bibitem[\protect\citeauthoryear{Vuckovic et al.}{2007}]{Vuckovic2007}
Vuckovic, M. et al. 2007, A\&A, 471, 605
\bibitem[\protect\citeauthoryear{Woolf \& Wallerstein}{2006}]{WW2006}
Woolf, V.~M. \&  Wallerstein, G., 2006, PASP, 118, 218


\end{thebibliography}
\end{document}